\documentclass[conference]{IEEEtran}
\IEEEoverridecommandlockouts
\usepackage{cite}
\usepackage{amsmath,amssymb,amsfonts}
\usepackage{algorithmic}
\usepackage{graphicx}
\usepackage{textcomp}
\usepackage[dvipsnames]{xcolor}
\usepackage{subfigure}
\usepackage{diagbox}
\usepackage{colortbl}
\usepackage{multirow}
\def\BibTeX{{\rm B\kern-.05em{\sc i\kern-.025em b}\kern-.08em
    T\kern-.1667em\lower.7ex\hbox{E}\kern-.125emX}}
\begin{document}

\title{Shared-kernel Wavelet Neural Networks for Poisson Image Reconstruction}

\author{Yuanhao Gong, Tan Tang, Qianyan Liu}

\maketitle

\begin{abstract}
The Laplacian operator transforms the image into its Laplacian field, which usually is sparse and satisfies a stable distribution. On the other hand, an image can be uniquely reconstructed from its Laplacian field via solving a Poisson equation with a proper boundary condition. Such uniqueness is mathematically guaranteed. Thanks to these properties, we propose to use the sparse Laplacian field to present the image. We first show that the Laplacian field is sparse and satisfies a stable distribution on hundreds images. Then, we show that the image can be accurately reconstruct from its Laplacian field. For the reconstruction task, we propose a shared-kernel wavelet neural network, which solves the Poisson equation and has three advantages. First, it has less than  {\bf 0.0002M} parameters, which is compact enough for most of devices. Second, it has linear computation complexity, leading to a real-time reconstruction. Third, it achieves higher accuracy than previous methods. Several numerical experiments are conducted to show the effectiveness and efficiency of the sparse Laplacian field and the proposed Poisson solver. The proposed method can be applied in a large range of applications such as image compression, low light enhancement, object tracking, etc.
\end{abstract}


\section{Introduction}
\label{sec:intro}
Images are fundamentally important for various applications, such as medical diagnosis, cell imaging, and visual navigation in robotics. In the medical imaging, high-resolution images from MRI and CT scans are crucial for detecting and monitoring diseases. In cell imaging, images play a pivotal role in understanding cellular processes and structures. Microscopy techniques, including fluorescence and electron microscopy, enable scientists to observe cells and their components in exquisite detail. Visual navigation in robotics relies heavily on image processing and computer vision. Robots use images to interpret their environment, identify objects, and navigate through spaces. This is particularly important in applications like autonomous vehicles, where quickly and accurately processing visual information is critical for safe operation.

Image representation is the research topic that studies how to represent the images from cameras, microscope or CT scan. In general, there are two types of image representation. One is the analog images, such as images from film cameras. The other is the digital images, such as images from CMOS cameras. In the past decades, thanks to the significant improvement in smart phones, the digital images are getting more and more popular. With the image resolution increasing, the image file size is also getting larger and larger, which becomes a critical issue for image representation.

Nowadays, many encoders are developed to represent the image, trying to reduce the file size. This task is called image compression. They usually transform the intensity image by discrete cosine transformation or wavelet transform, obtaining a sparse representation. Although these methods are popular, the sparsity is varying because the image intensity distribution mainly depends on the image content. 

\begin{figure}
	\centering
	\includegraphics[width=\linewidth]{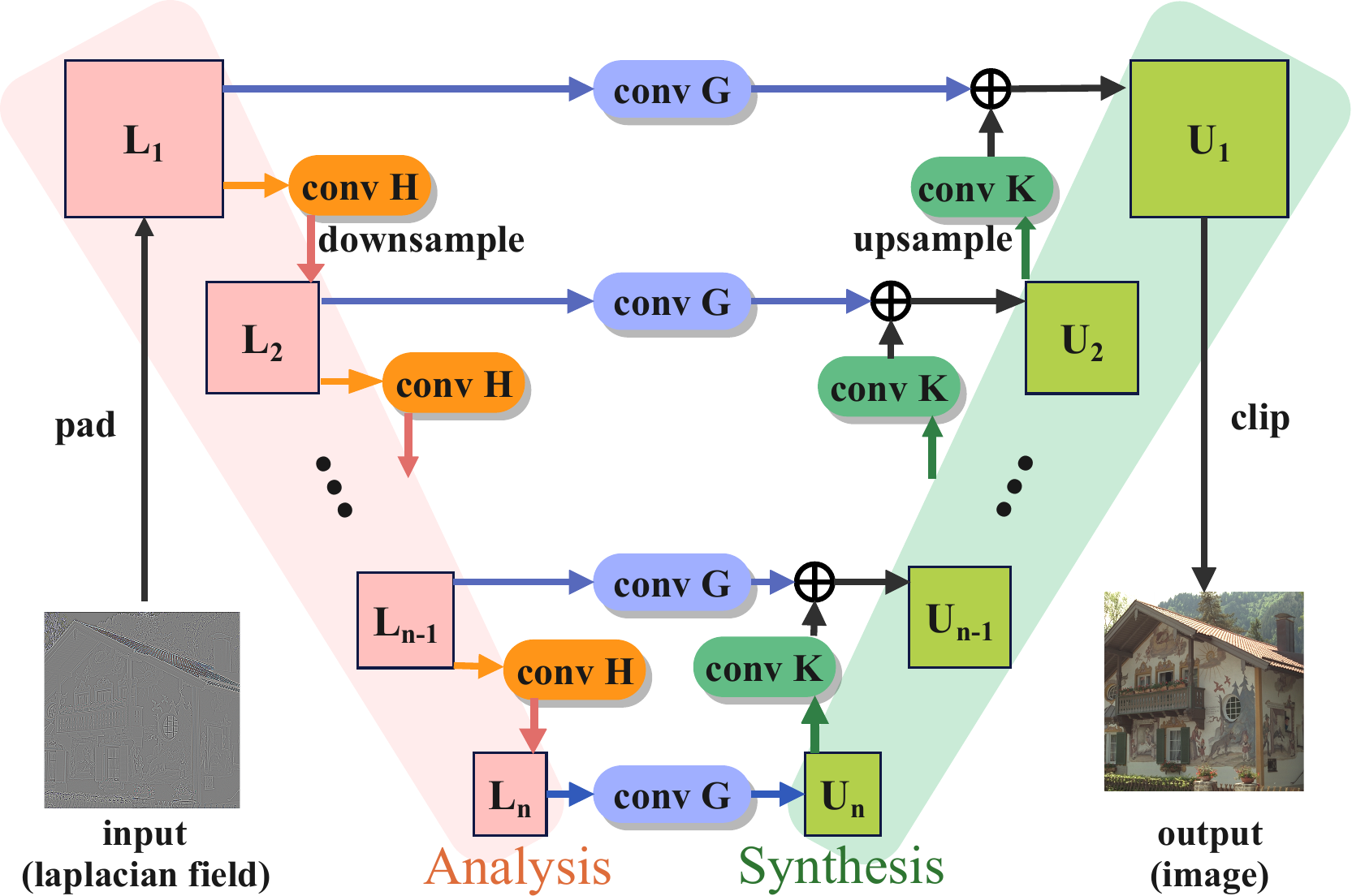}
	\caption{The pipeline of the proposed wavelet guided convolution neural network to reconstruct the image from its sparse Laplacian field. The shaded {\color{Salmon}red stream }is decomposition process (analysis in the wavelet transform) for the input Laplacian field $L$ while the shaded {\color{Green} green stream} is the reconstruction process (synthesis in the wavelet transform) for the output $U$. In this network, there is {\bf no fixed wavelet basis}. Instead, the convolution kernels (H, G and K) are {\bf learned} and {\bf shared}, leading to less than {\bf 0.0002M} parameters regardless the input resolution. Moreover, it has linear computation complexity with respect to the input pixels, leading to a high computation performance. }
	\label{fig:pipe}
\end{figure}

Luckily, the gradient domain~\cite{Strong:1996} and the Laplacian domain images~\cite{gong:phd} are sparse and statistically stable. The gradient is the first order derivative of the image, which indicates the image edges and contrast. The Laplacian domain is the second order derivative of the image, which can capture the image edges as well. Although they are different, they have a close relationship as we will explain the the following subsections.
\subsection{Gradient Field}
Gradient domain image processing has a long history~\cite{fattal2002gradient,Fei:2015}. The large gradient usually indicates the image edges, which contain more visual information than the flat regions. In the gradient domain image processing, the pipeline usually contains three steps. First, the gradient of the original input image is computed. The gradient operator is discretized by finite difference schemes. Second, the gradient is manipulated to obtain a new gradient field, which contains the preferred target properties. Third, the output image is reconstructed from the new gradient field by solving a Poisson equation.

More specifically, if the target gradient field is $\vec{G}=(g_x,g_y)$, we try to find the $u(\vec{x})$ minimizing the following model
\begin{equation}
	\label{eq:loss}
	{\cal L}(u(\vec{x}))=\frac{1}{2} \|\nabla u(\vec{x}) - (g_x,g_y)\|^2_2\,,
\end{equation} where $\nabla$ is the gradient operator.

The optimal condition of the loss function~\eqref{eq:loss} is 
\begin{equation}
	\frac{\partial {\cal L}}{\partial u}=\nabla^T\nabla u-\nabla^T(g_x,g_y)=0\,,
\end{equation} leading to
\begin{equation}
	\label{eq:lap0}
	\Delta u=\nabla^T(g_x,g_y)\,,
\end{equation}
which is a standard Poisson equation.

Be aware the the gradient of the resulting $u(\vec{x})$ might not be exactly the target gradient field $(g_x, g_y)$, but is close to the target in the $\ell_2$ norm sense as shown in Eq.~\eqref{eq:loss}. Therefore, the accurate estimation of the gradient fields is fundamentally important in such methods.

Gradient domain image processing has been used in various applications such as HDR~\cite{fattal2002gradient}, image enhancement~\cite{Paris2011,gong:gdp} and registration~\cite{Duan2023}. In all these applications, the core task is to efficiently and effectively solve the Poisson equation.

\subsection{Laplacian Field}
We notice that the gradient domain image processing essentially becomes a Poisson equation with a Laplacian field. Therefore, we propose to directly work in the Laplacian field. 
More specifically,
the Eq.~\eqref{eq:lap0} can be generalized as
\begin{equation}
	\label{eq:lap}
	\Delta u(\vec{x})=L(\vec{x})\,,
\end{equation}where $\Delta\equiv\nabla^T\nabla=\frac{\partial^2}{\partial x^2}+\frac{\partial^2}{\partial y^2}$ is the Laplacian operator and $L$ is the observed Laplacian field. The Poisson equation frequently appears in mathematics, physics and computer science. It has been studied a lot. Its numerical solvers will be discussed in the later sections.

Although the gradient field has a close relationship with the Laplacian field, there are three main differences between the gradient field and the Laplacian field.
\begin{itemize}
	\item Laplacian field is from a {\bf second} order derivative while the gradient field is from the {\bf first} order derivative. As a result, the Laplacian field is theoretically {\bf more  sparse}.
	\item Laplacian field is a {\bf scalar} field while the gradient field is a {\bf vector} field. Therefore, the Laplacian field needs only half storage and is more computationally efficient.
	\item Laplacian field has a unique solution via solving the Poisson equation while the gradient field might not be integrable ($g_x$ and $g_y$ might not consist).
\end{itemize}

\subsection{Our Motivation and Contributions}
We propose to use the Laplacian fields for image representation for three reasons. First, the Laplacian field is sparse, which indicates the computation and storage efficiency. Second, the Laplacian field satisfies a stable distribution. Third, the Laplacian field can be used to accurately recover the original image via solving a Poisson equation. And there are many efficient Poisson solvers available. Our contributions are
\begin{itemize}
  \item We propose to use Laplacian fields to encode images.
  \item We numerically confirm the sparsity and stability of the Laplacian fields.
  \item We propose a neural network Poisson solver that can reconstruct the image from  its Laplacian field.
  \item Several experiments are conducted to show the efficiency and the effectiveness of the proposed method.
\end{itemize}
\section{Laplacian Fields}
In this section, we show several advantages of the Laplacian fields. More specifically, the Laplacian field is sparse, satisfies a stable distribution, and has a unique solution via solving the Poisson equation. Such uniqueness is mathematically guaranteed, which provides the theoretical root for our proposed method. With these advantages, the sparse Laplacian fields are preferred as image presentation.

\begin{figure}
	\centering
	\subfigure[image intensity]{\includegraphics[width=0.3\linewidth]{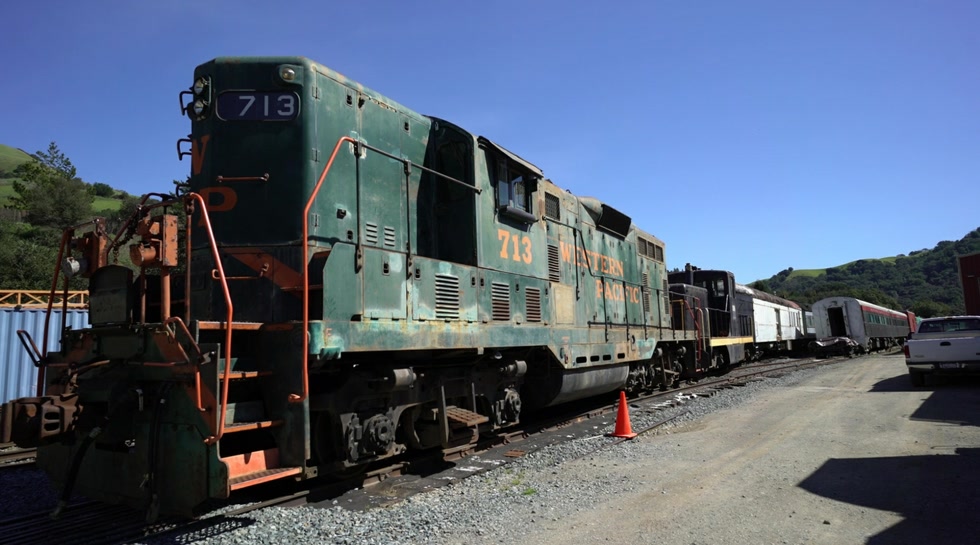}}
	\subfigure[Gradient X]{\includegraphics[width=0.3\linewidth]{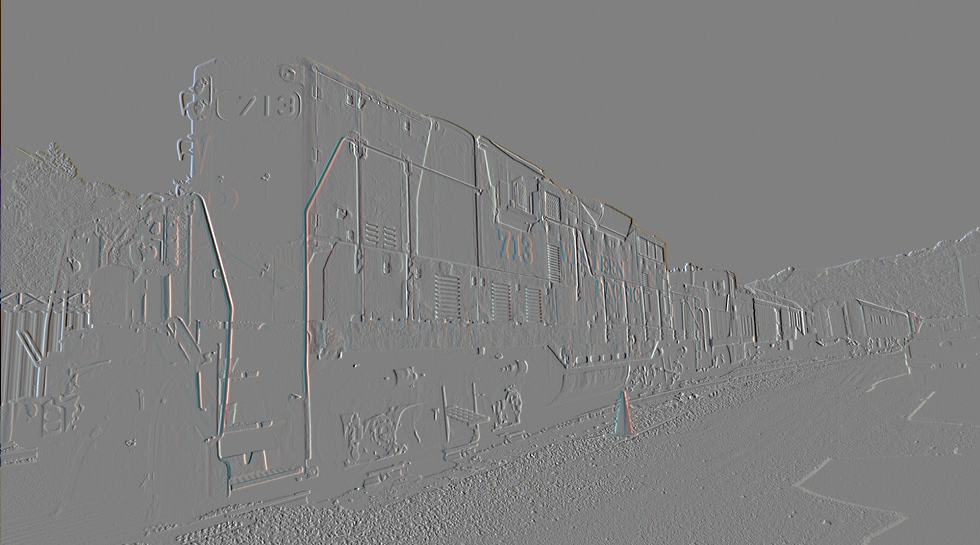}}
	\subfigure[Laplacian field]{\includegraphics[width=0.3\linewidth]{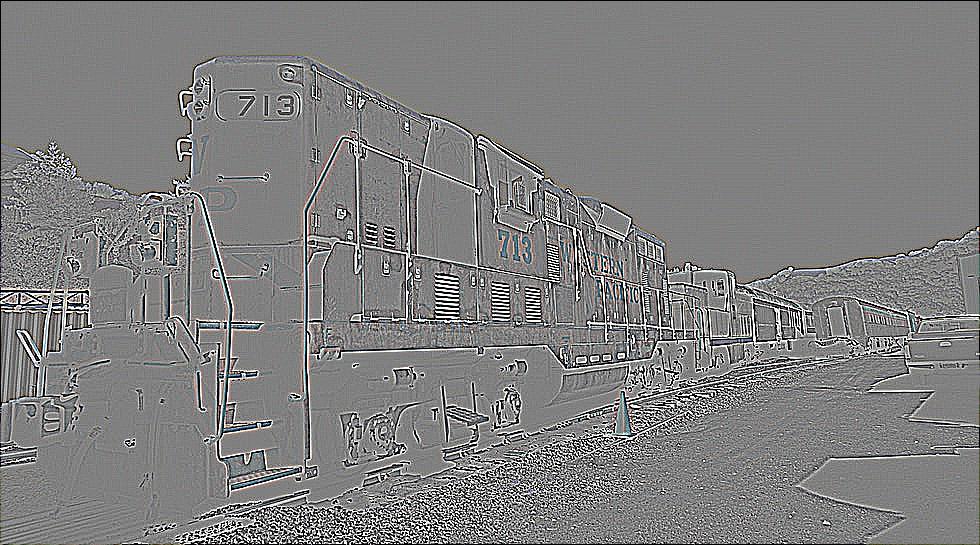}}
	\subfigure[intensity distribu.]{\includegraphics[width=0.3\linewidth]{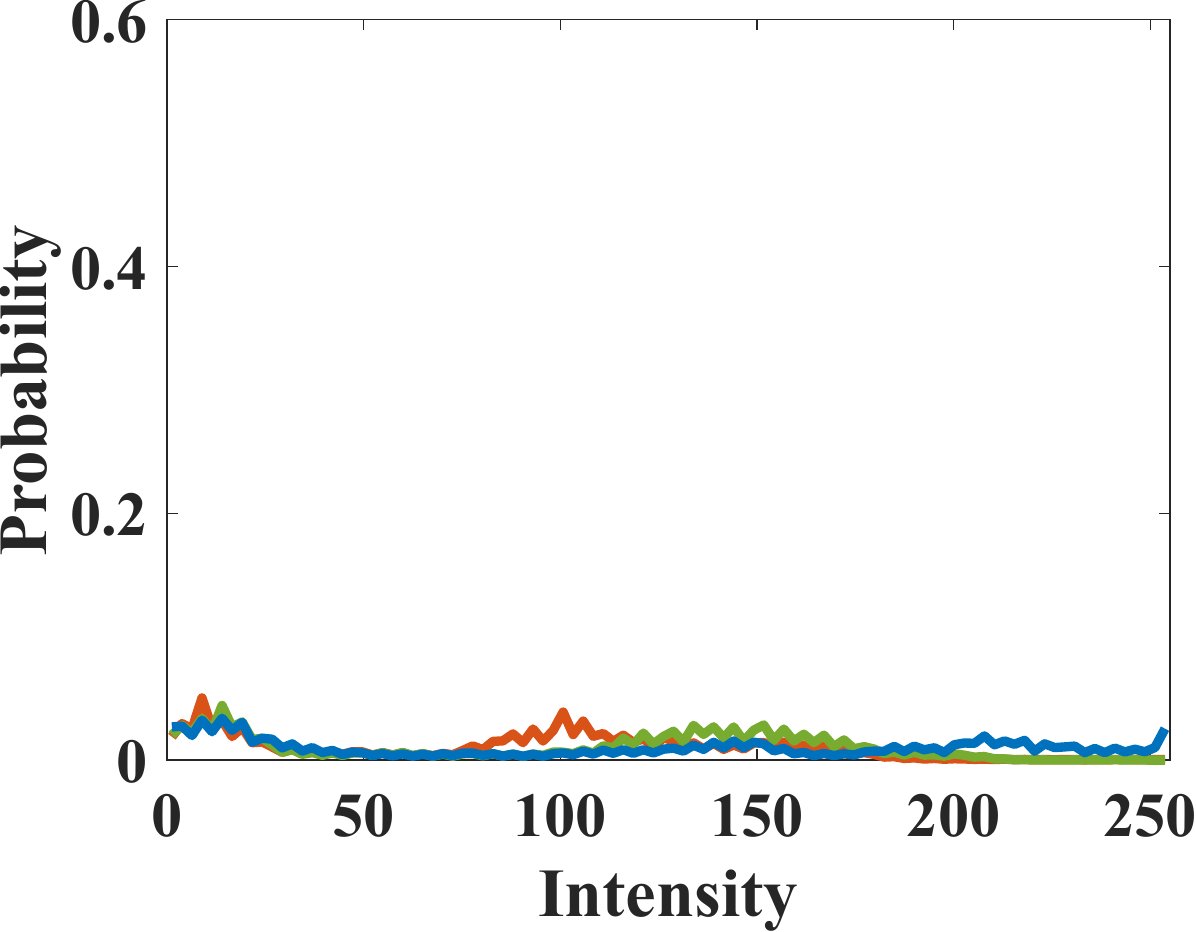}}
	\subfigure[GradientX distribu.]{\includegraphics[width=0.3\linewidth]{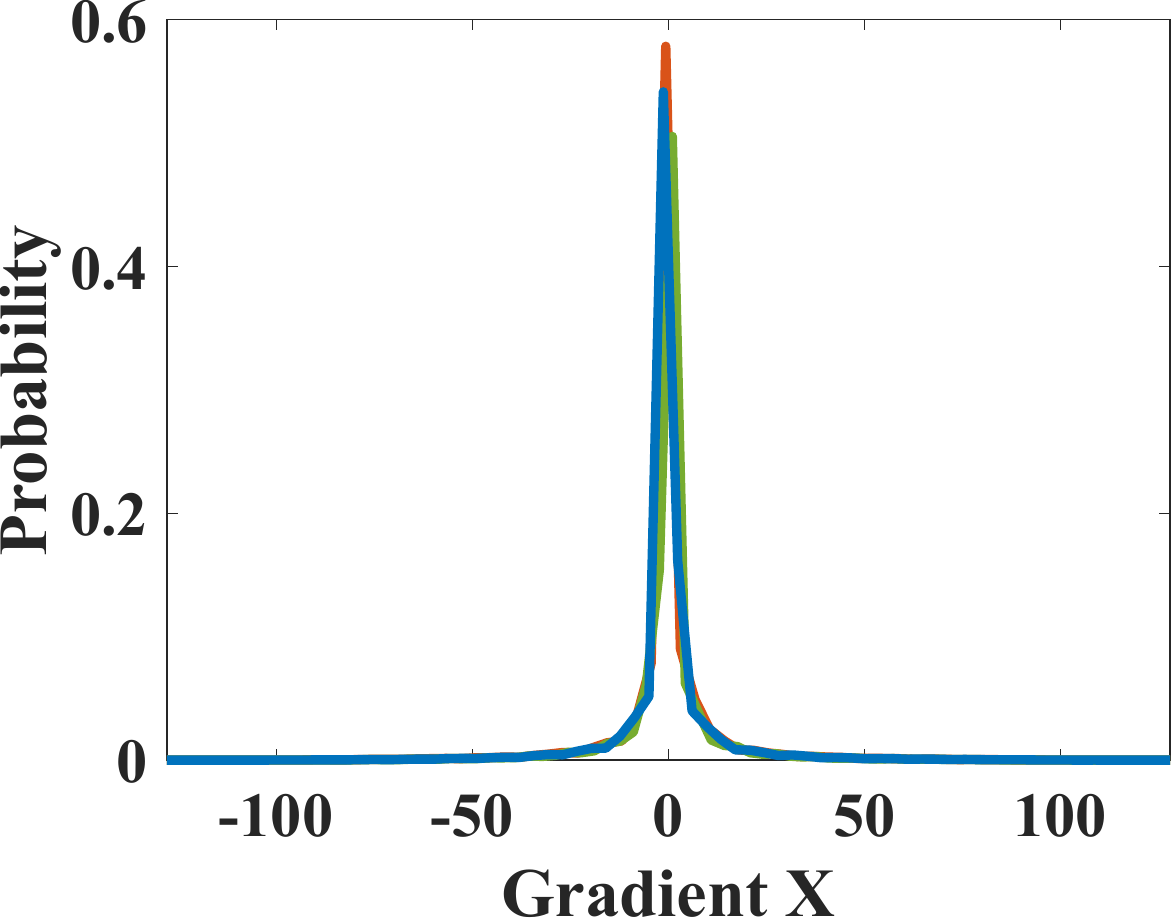}}
	\subfigure[Laplacian distribu.]{\includegraphics[width=0.3\linewidth]{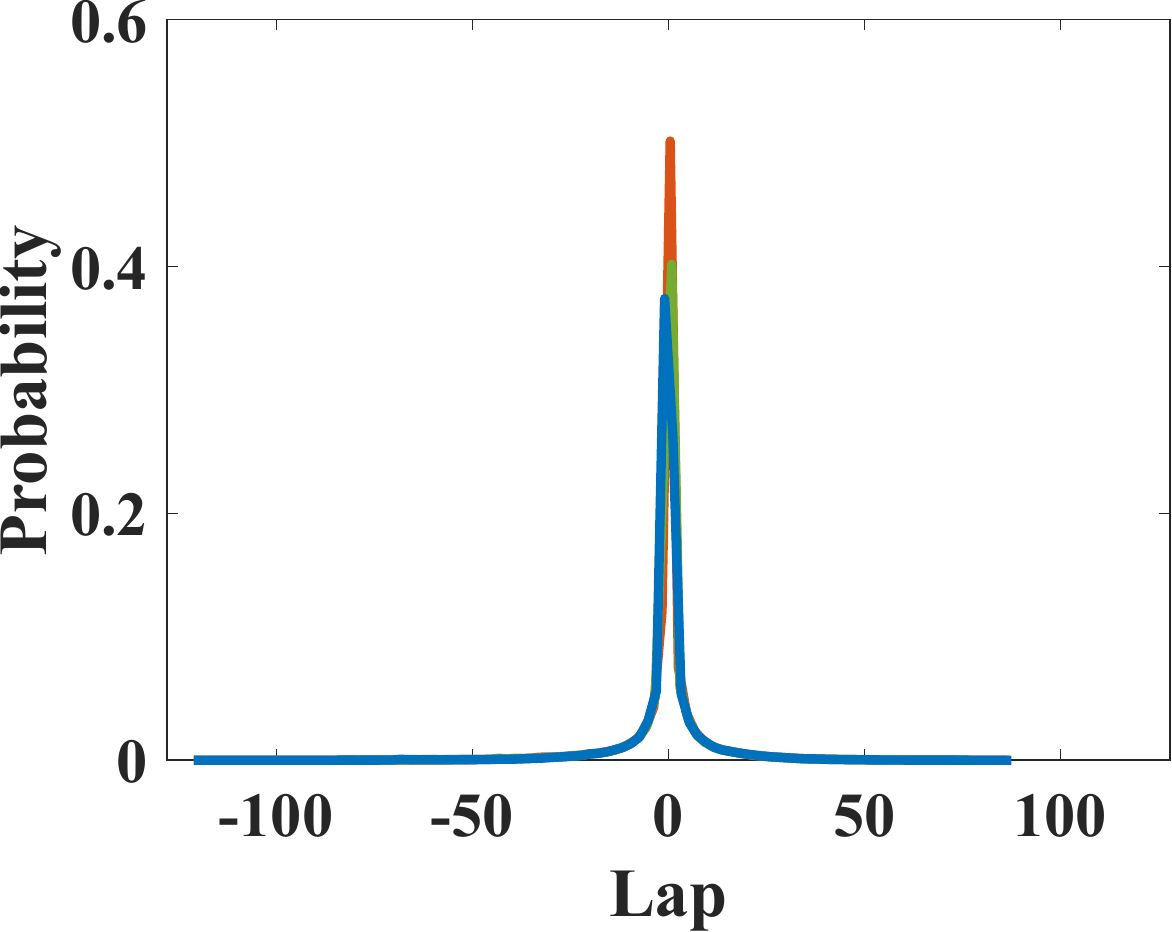}}
		\subfigure[intensity CDF]{\includegraphics[width=0.3\linewidth]{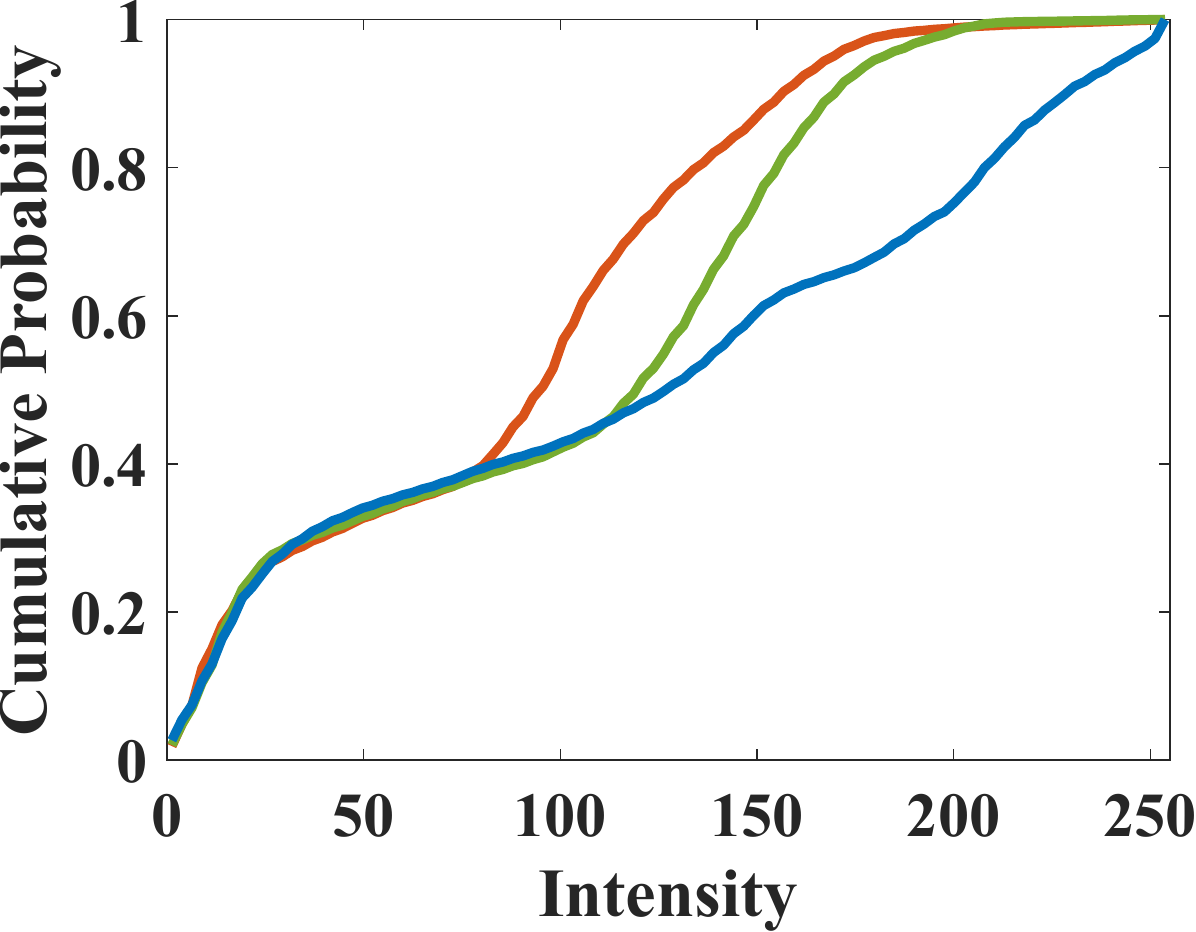}}
		\subfigure[Gradient X CDF]{\includegraphics[width=0.3\linewidth]{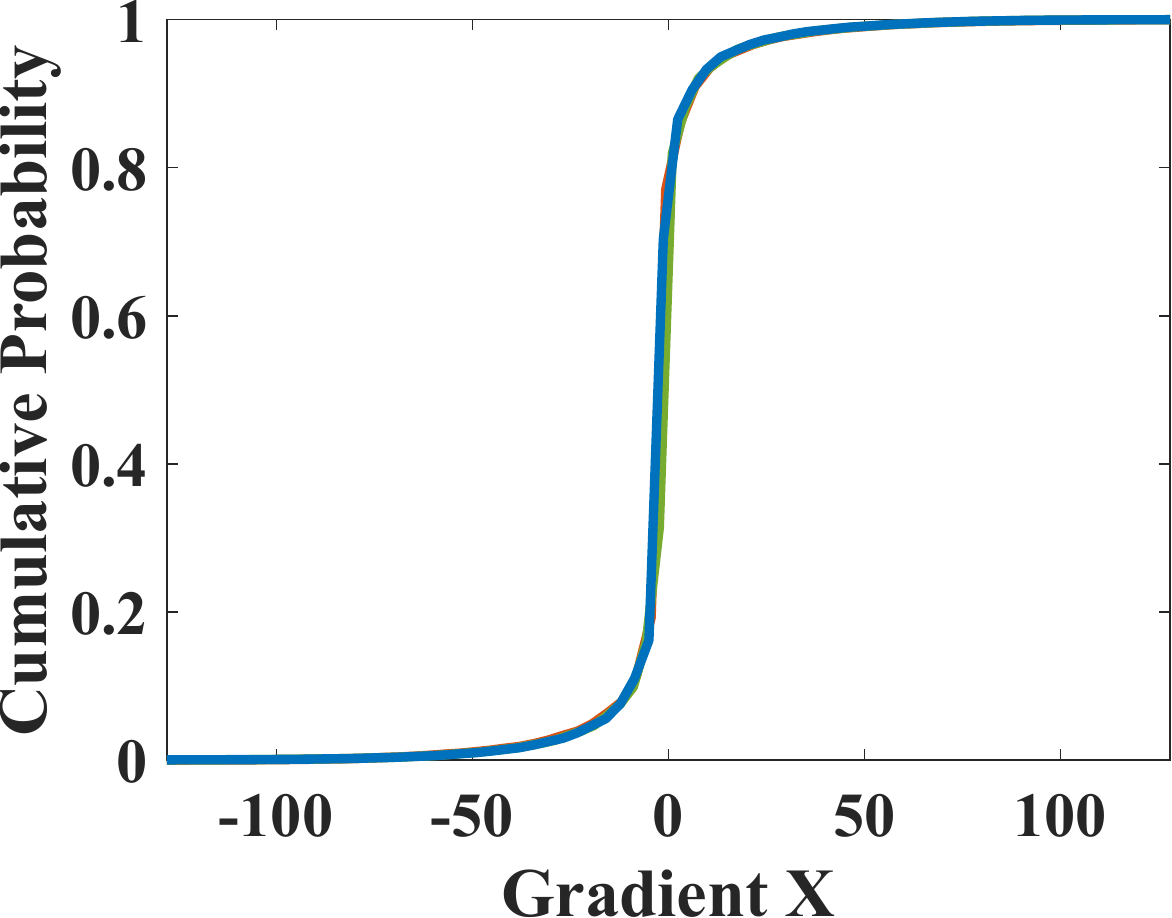}}
	\subfigure[Laplacian CDF]{\includegraphics[width=0.3\linewidth]{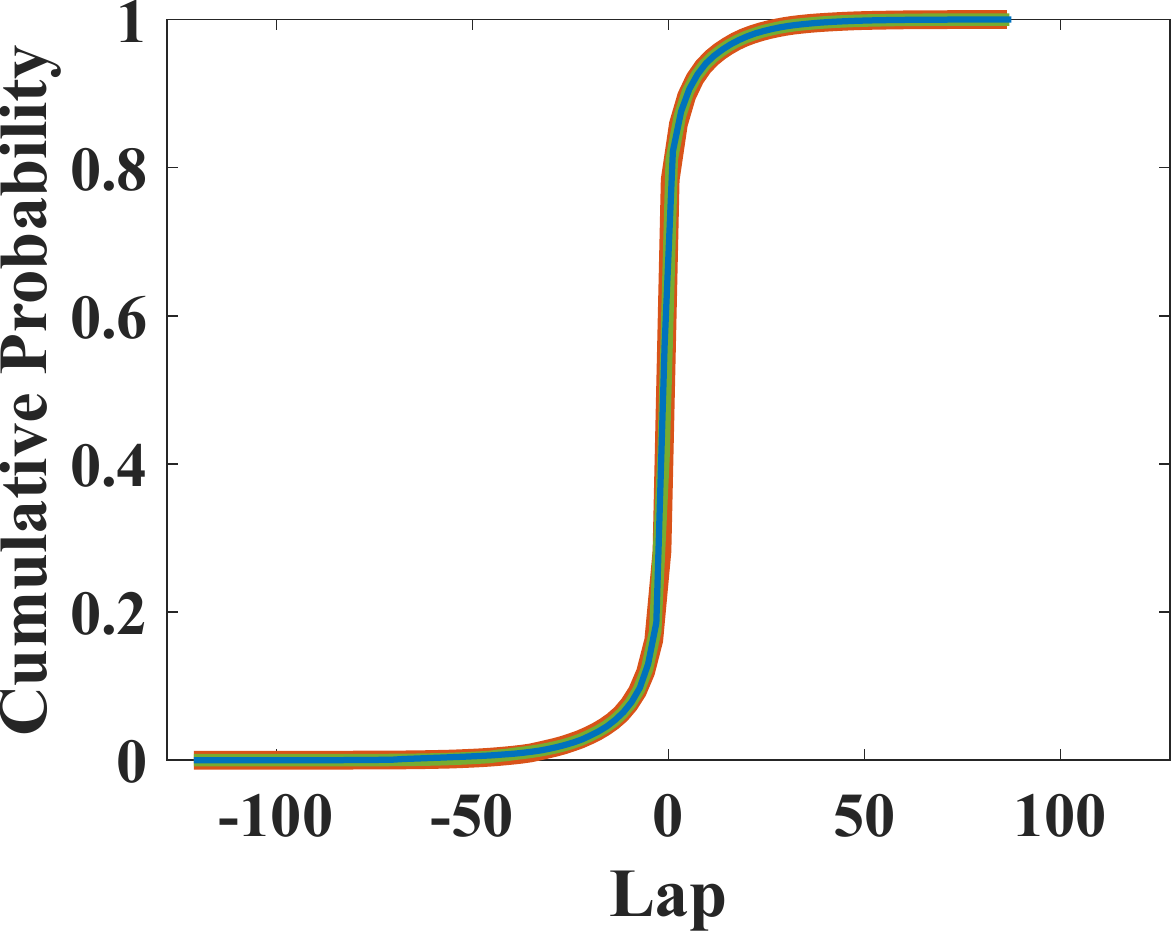}}
	\caption{For the input image (a), its intensity satisfies almost a uniform distribution as shown in (d) and (g). In contrast, its gradient and Laplacian fields are much sparser as shown in (e) and (f). Be aware that gradient fields have x and y two fields while Laplacian field has only one. The red, green and blue colors indicate different color channels, respectively.}
	\label{fig:1}
\end{figure}

\subsection{Sparsity}
First of all, the Laplacian field is sparse~\cite{Gao2013}. One example is shown in Fig.~\ref{fig:1}. One image is shown in Fig.~\ref{fig:1}(a) and its intensity distribution is shown in Fig.~\ref{fig:1}(c), along with its cumulative density function in Fig.~\ref{fig:1}(e). Its Laplacian filed is shown in Fig.~\ref{fig:1}(b) and its intensity distribution is shown in Fig.~\ref{fig:1}(d), along with its cumulative density function in Fig.~\ref{fig:1}(f). We can conclude that the Laplacian field is much sparser (more items around zero). This fact is not a coincidence, but generic for most images.

\begin{figure}
	\centering
	\subfigure[BSDS500]{\includegraphics[width=0.45\linewidth]{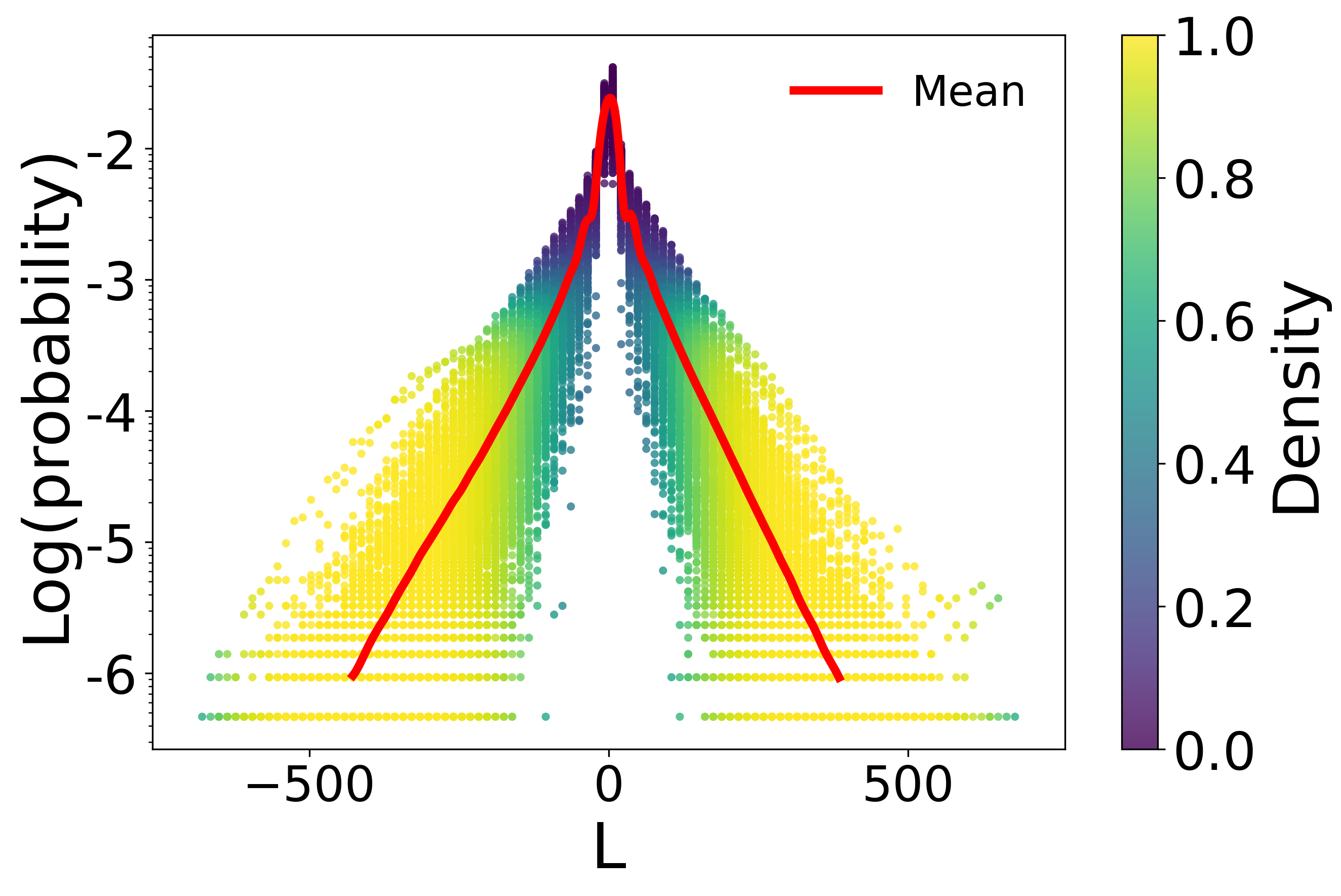}}~~
	\subfigure[DIV2K]{\includegraphics[width=0.45\linewidth]{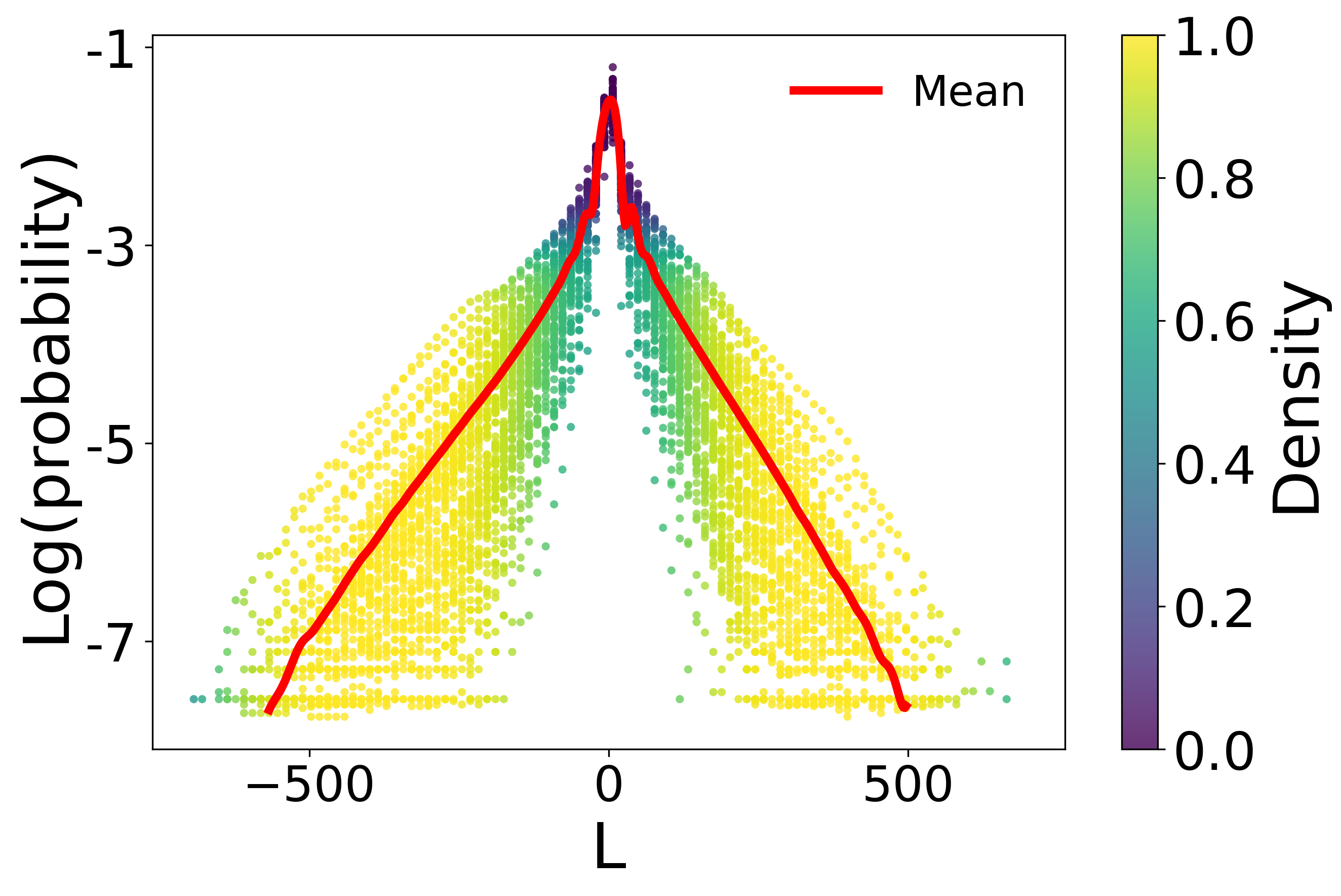}}
	\caption{The statistics of the Laplacian fields from BSDS500 data set (a) and the DIV2K data set. BSDS500 contains 500 images with resolution $481\times 321$. DIV2K contains 202 images with different resolutions.}
	\label{fig:2}
\end{figure}

To numerically confirm the sparsity of the Laplacian fields, we show the distribution of Laplacian fields from 500 natural images in BSDS500 data set with low resolutions in Fig.~\ref{fig:2}(a). The same statistics of higher resolution images in data set DIV2K is shown in Fig.~\ref{fig:2}(b). Although these images have different content and resolutions, their Laplacian fields are sparse. And the average distribution is shown as the red line in each figure. The red lines roughly satisfy an $\ell_1$ norm in the $\log$ scale, indicating a Laplacian Distribution. These results confirm the sparsity of the Laplacian fields.

\if false
We can also use a threshold parameter $T$ to manually force the sparsity of the Laplacian field. For example, we count the Laplacian field $|L|>T$ in Fig.~\ref{fig:1} (b) and shows the percentage with respect to the $T$ in Fig.~\ref{fig:sparse}. When $T=4.5$, the half of all pixels are nonzero.
\begin{figure}[!b]
	\centering
	\includegraphics[width=0.5\linewidth]{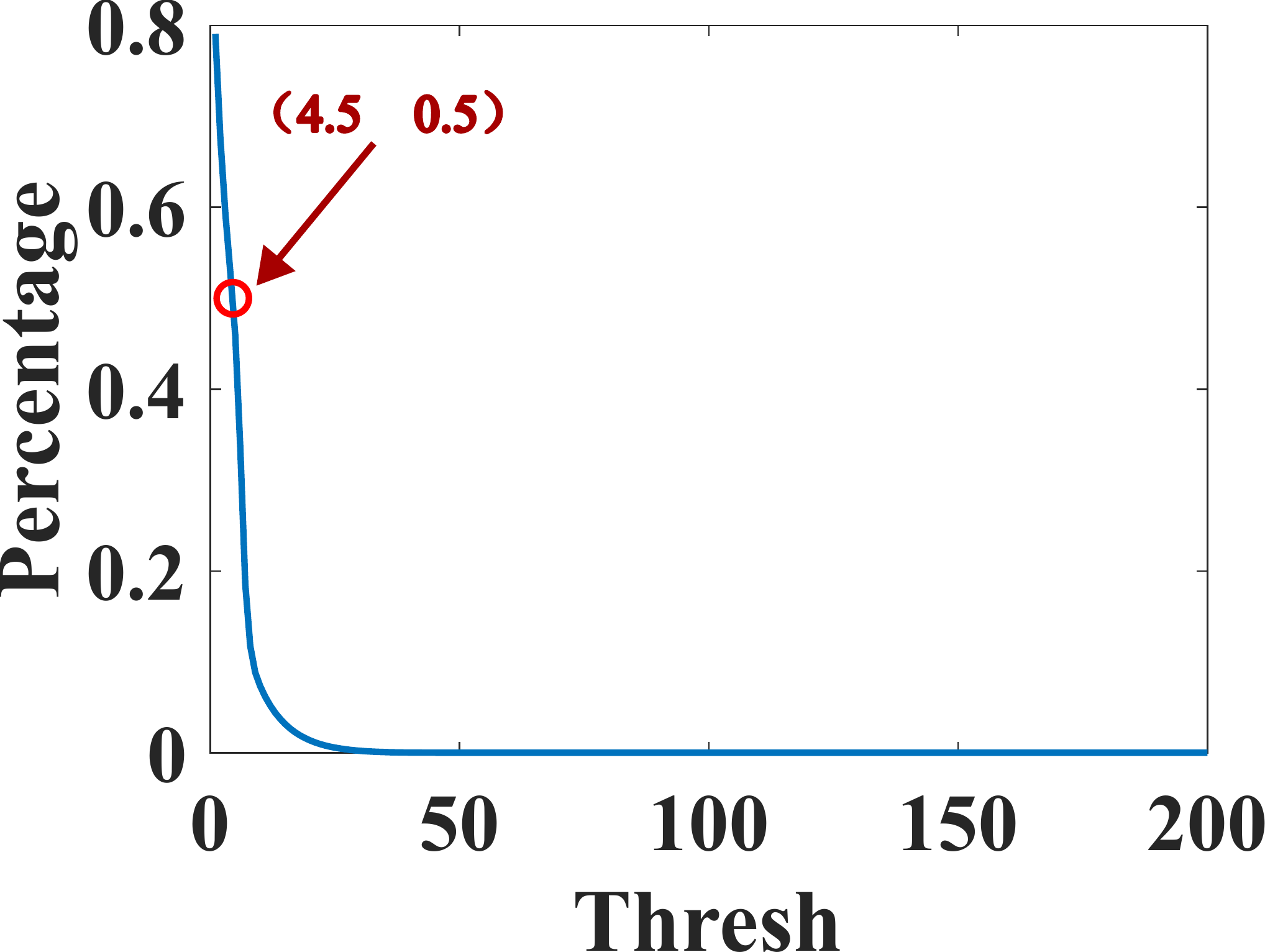}
	\caption{The sparsity with respect to the threshold parameter $T$}
	\label{fig:sparse}
\end{figure}
\fi
\subsection{Stability}
Second, the Laplacian fields from different images satisfy a stable distribution. It is well-known that the distributions (histogram) of image intensities are varying dramatically, depending on the image content. In contrast, the gradient distribution is more stable~\cite{gong:gdp}. As the result, the Laplacian field is also stable. As shown in Fig.~\ref{fig:2}, each distribution is varying around the average distribution (the red line).

\subsection{Uniqueness}
Finally, the Poisson equation with the given Laplacian field, Eq.~\eqref{eq:lap}, has a unique solution with a proper boundary condition. More specifically, we use the Dirichlet boundary condition:
\begin{equation}
	\left\{
	\begin{array}{ll}
		\Delta u(\vec{x})=L(\vec{x}),  &\vec{x}\in \Omega \\
		u(\vec{x})=0,  &\vec{x}\in\partial \Omega
	\end{array}\right.\,,
\end{equation} where the $\Omega$ is the imaging domain and $\partial \Omega$ indicates the boundary of the imaging domain. Such uniqueness is theoretically guaranteed from mathematics~\cite{Jackson1998}, named Uniqueness Theorem for Poisson equation..

Thanks to the uniqueness of the Poisson equation and the sparsity of the Laplacian field, the linear system can be efficiently solved. The solving algorithms will be explained in the following section~\ref{sec:nn}. 
\subsection{Discrete Laplacian Operator}
The Laplacian operator $\Delta$ can be discretized via different finite different schemes. In the image processing community, the following convolution stencils are popular~\cite{Gong2019,Gong2022}
\begin{alignat}{2}
	k_0 &= \left[ \begin{array}{ccc} 0 & 1 & 0 \\
		1 & -4 & 1 \\
		0 & 1 & 0 
	\end{array} \right]\!,\,&
	k_1 &= \left[ \begin{array}{ccc} \frac{1}{2} & \frac{1}{2} & \frac{1}{2} \\
		\frac{1}{2} & -4 & \frac{1}{2} \\
		\frac{1}{2} & \frac{1}{2} & \frac{1}{2} 
	\end{array} \right]\,,
\end{alignat}
	\begin{alignat}{2}
	k_2 &= \left[ \begin{array}{ccc} -\frac{1}{4} & \frac{5}{4} & -\frac{1}{4} \\
		\frac{5}{4} & -4 & \frac{5}{4} \\
		-\frac{1}{4} & \frac{5}{4} & -\frac{1}{4} 
	\end{array} \right]\!,\,&
	k_3 &= \left[ \begin{array}{ccc} \frac{1}{3} & \frac{2}{3} & \frac{1}{3} \\
		\frac{2}{3} & -4 & \frac{2}{3} \\
		\frac{1}{3} & \frac{2}{3} & \frac{1}{3} 
	\end{array} \right].
	\label{eq:kernel}
\end{alignat}

In this paper, we choose the kernel $k_0$ to approximate the Laplacian operator. First, this kernel is derived from the classical finite different schemes and has been numerically validated in various applications. Second, the negative of this kernel is separable, indicating a faster computation. Third, its computation only depends on four neighbors while other kernels rely on eight neighbors. This indicates $k_0$ is more compact (local) than others. From geometric point of view, the smaller support region, the better geometric description.
\section{Shared-kernel Wavelet Neural Network}
\label{sec:nn}
Assuming the Laplacian field $L(\vec{x})$ is given, the original image $U(\vec{x})$ can be recovered by solving a Poisson equation. The Poisson equation is common in various scientific research and engineering applications. It has been studied a lot, leading to many numerical solvers. We give a brief introduction about previous solvers before explaining our method.
\subsection{Previous Solving Methods}
In general, there are two types of Poisson solvers. One is the traditional methods which usually depend on the finite difference schemes. The other is the deep learning methods that usually are data-driven. We give them a brief introduction before explaining our solving method. 
\subsubsection{Traditional Solvers}
In many finite element methods, the Laplacian operator is discretized by finite difference schemes
\begin{equation}
	{
	\Delta u\approx u_{i-1,j}+u_{i+1,j}+u_{i,j-1}+u_{i,j+1}-4u_{i,j}\,,}
\end{equation}
which leads to a diagonal-dominant matrix which usually is semi negative definite. 
Then, iterative methods such as  Jacobian or Gauss-Seidel iteration can be used to solve the linear equation and find the solution for a given tolerance.

\begin{table*}
	\caption{Summary of traditional Laplace solvers for $N$ samples. } 
	\label{table:Poisson} 
	\centering  
	\begin{tabular}{c|cccccccc} 
		\hline\hline
		Solver  & Cholesky & Jacobian&	GaussSeidel & SOR & {FFT} &Multigrid &  { Wavelet}& {\bf Ours} \\
		\hline
		Type & direct & iterative & iterative	&	iterative& 	 direct &iterative & direct&	{direct}\\
		\hline
		Complexity& \cellcolor{orange!50}${ O}(N^3)$ & \cellcolor{orange!30} ${ O}(N^2)$ & \cellcolor{orange!30} ${ O}(N^2)$ &  \cellcolor{orange!15}${ O}(N^{3/2})$&  \cellcolor{orange!7}${ O}(N\log N)$ & ${ O}(N)$&	${ O}(N)$& {\bf ${ O}(N)$}\\
		\hline
	\end{tabular}
\end{table*}

A multiscale strategy such as multigrid can be adopted to further accelerate the computation. The multigrid methods solve the equation at coarse level and interpolate in the finer level. By repeating this procedure, the final result is reached at the finest scale. Although these methods are iterative, they have linear computational complexity thanks to the multiscale.

Instead of performing the iteration, the Discrete Sine Transform (DST) can be used to solve the Poisson equation. For example, the second order derivative is represented by a convolution with a kernel $[1,-2,1]$ in 1D. Using the relationship $\sin((i-1)x)+\sin((i+1)x)=2\cos(x)\sin(ix)$, we have
\begin{equation}
	\begin{split}
		-\Delta\sin(ix)&=	2\sin(ix)-\sin((i-1)x)-\sin((i+1)x)\\&=2(1-\cos(x))\sin(ix)\\&
		=4\sin^2(\frac{x}{2})\sin(ix)\,.
	\end{split}
\end{equation}
Therefore, $\sin(ix)$ is an eigenfunction of $-\Delta$ and we can use the DST of the Laplacian field $L$ to directly find the solution $U$ without iterations. Such methods are called direct methods or spectral methods. The DST or FFT based methods have $O(N\log(N)) $ computation complexity.

Another direct method is the wavelet solver~\cite{Jia2006,Farbman2011}, which has linear complexity. These methods decompose the Laplacian fields into wavelet coefficients. After solving the equation in each scale, the inverse wavelet transform is performed to find the solution. One important work in this type of methods is the Cubic Hermite Spline Wavelet~\cite{gong:phd}, which has a bounded condition number $\frac{15}{4}$. This is the only solver that has a bounded condition number for solving Poisson equation. 

These traditional methods are summarized in Table~\ref*{table:Poisson}, along with their solving type and computation complexity. As can be told, the linear computation complexity is the lowest among all solvers. And our method can also achieve this linear complexity, leading to a real-time performance. 
\subsubsection{Neural Network Poisson Solvers}
With the development of deep learning, many neural network solvers for Poisson equation have been studied such as~\cite{Tang2017,Shan2020,GUOZHENG2023, Lu2024,Yu2024,Brunton2024,Hu2024}. Very recently, the deep learning method can also deal with PDEs on different geometries~\cite{Yin2024}. In general, these methods are data-driven and thus can achieve a higher accuracy. 

However, previous deep neural networks have three limitations, the large model size, the high computation cost, and weak generalization. First, they have a large number of parameters, hampering their application in practice. Reducing the parameter number will reduce the accuracy accordingly. Second, they have to be performed on an advanced GPU with a large computer memory. For example, the pre-processing step can use two hours for a $256\times 256$ image. Third, these networks can not be easily extended for different resolution images. In their settings, the resolution of inference images must be exactly the same as the training data set.
\subsection{Our Shared-kernel Wavelet Neural Network}
Inspired by previous wavelet solvers and the deep learning methods, in this paper, we propose a novel neural network Poisson solver that has advantages from both. The pipeline of our network is shown in Fig.~\ref{fig:pipe}, where we follow the wavelet philosophy but use data-driven basis functions. The shaded red and green regions indicate the decomposition (analysis) and reconstruction (synthesis) stream, respectively. 

During the decomposition, the input $L_1$ is convoluted with a kernel $H$ (corresponding to the low-pass filter in the wavelet) and then down sampled to reach the next scale level. This procedure is repeated until the final scale $L_n$ is reached. Then, $L_n$ is convoluted with a kernel $G$, leading to a reconstruction $U_n$. During the reconstruction process, $U_n$ is up-sampled and convoluted with a kernel $K$. At the same time, $L_i$ is convoluted with a kernel $G$ (corresponding to the high pass filter in the wavelet) to reduce the error during the reconstruction stream. 

Be aware that the learned convolution kernels $H, G, K$ are shared crossing all the scales. With such design, these learned kernels are independent from the scale. Moreover, the shared kernels can reduce the number of learnable parameters, leading to efficient training process. 

Thanks to the pure convolution design, the proposed network can be trained on any resolution images but applied on other resolution images. This makes our method have better generalization than other neural networks that have fixed the image resolution.

The proposed network has advantages from the traditional methods~\cite{Gong2023} and the deep learning approaches~\cite{Wei2024}. First, our network follows the wavelet fashion which naturally inherits the multiscale property. Second, the convolution kernels are learned from the training data. Therefore, it has data-driven property from the deep learning. More specifically,
\begin{itemize}
	\item The proposed network is multiscale, similar to the wavelet method.
	\item The filters in the proposed network are data-driven.
	\item The proposed network has only {\bf 177} parameters.
	\item  The trained network can work on any resolution images.
\end{itemize}

We use the standard $\ell_2$ norm as the loss function
\begin{equation}
	{\cal L}(U(\vec{x}))=\frac{\frac{1}{2}\|U(\vec{x})-GT(\vec{x})\|_2^2}{\int\mathrm{d}\vec{x}}\,,
\end{equation} where $U=WCNN(L)$ is the output of the network with input $L$ and $GT$ is the ground truth. For the data set, we further average of the sample axis and the color channel axis.

In general, we use the batch size =32, the learning rate $=10^{-5}$ and the epoch as 2000. 
\subsection{Ablation Study}
In this section, the ablation study is carried out to confirm the proposed network. More specifically, we investigate the image resolution, filter kernel size, and the scene adaptability. 
\subsubsection{Data Resolution}
In this work, we use two different resolutions images to train the network. One is the BSDS500 data set with resolution $481\times321$. We randomly sample 6400 image patches with resolution $256\times 256$ and augment these patches with rotation and mirror flip. We then sample 1600 image patches with the same resolution and augmentation as the test data set. This is a low-resolution data set. 

To show the performance of the proposed network, we also use the DIV2K data set with the resolution $2040\times 1356$. We sampled 1600 patches with resolution $1024\times 1024$ as the training data and 200 patches as the testing data. This is a high-resolution data set.

The results are shown in Table~\ref{tab:loss_comparison}. We can conclude that the classical methods such as wavelet~\cite{Farbman2011} work better on high resolution images (DIV2K). In contrast, our method works better on low resolution images (BSDS500). In general, our method works better than previous direct solvers.
\begin{table}[htbp]
	\caption{MSE loss with different kernel size on two data sets}
	\begin{center}
		\begin{tabular}{|c|c|c|}
			\hline
				\rowcolor{gray!30} Methods & \textbf{DIV2K}($1024\times1024$) & \textbf{BSDS500}($256\times 256$) \\
			\hline
			wavelet~\cite{Farbman2011} & 5.03919 & 6.32709 \\
			\hline
		\rowcolor{blue!15}	ours 5$\times$5 & 4.90728 & 4.26689 \\
		\rowcolor{blue!10}	ours 7$\times$7 & 2.76564 & 1.21060 \\
		\rowcolor{blue!4}	ours 9$\times$9 & 0.86227 & 0.40010 \\
			ours 11$\times$11 & 0.60981 & 0.33186 \\
			\hline
		\end{tabular}
		\label{tab:loss_comparison}
	\end{center}
\end{table}
\subsubsection{ Kernels}
We studied different convolution kernel size. And the results are shown in Table~\ref{tab:loss_comparison}. In general, a larger kernel size leads to a higher accuracy. Since the computation of the network is not impacted by the convolution kernel size, a large kernel size is suggested if the high accuracy is required.

To analyze the learned kernels, $5\times5$ kernel $H$, $3\times3$ kernel $G$, and $5\times5$ kernel $K$ are explicitly given in Table~\ref{tab:H_matrix}. These kernels do not have the zero-mean as required in wavelets. 
\begin{table}[htbp]
	\caption{The Learned kernel $H$ (top) , $G$ (middle) and $K$ (bottom)}
	\centering
	\begin{tabular}{|c|c|c|c|c|}
		\hline
		0.020 & 0.072 & 0.106 & 0.077 & 0.019 \\
		\hline
		0.072 & 0.245& 0.348 & 0.249 & 0.072 \\
		\hline
		0.107 & 0.348 & 0.495 & 0.350 & 0.108 \\
		\hline
		0.077 & 0.249 & 0.350 & 0.254 & 0.076 \\
		\hline
		0.019 & 0.072& 0.107 & 0.077 & 0.016\\
		\hline
	\end{tabular}
	\label{tab:H_matrix}\\
	\vspace{2mm}
	\begin{tabular}{|c|c|c|}
		\hline
		0.029 & 0.087 & 0.028 \\
		\hline
		0.087 & 0.311 & 0.087 \\
		\hline
		0.028 & 0.087 & 0.028 \\
		\hline
	\end{tabular}
	\label{tab:G_matrix}
	\vspace{2mm}
	\begin{tabular}{|c|c|c|c|c|}
		\hline
		-0.023 & 0.091 & 0.188 & 0.083 & -0.020 \\
		\hline
		0.090  & 0.262 & 0.331 & 0.252 & 0.094 \\
		\hline
		0.187  & 0.331 & 0.340 & 0.327 & 0.189  \\
		\hline
		0.082  & 0.252 & 0.328 & 0.246 & 0.087  \\
		\hline
		-0.020 & 0.094 & 0.190 & 0.088 & -0.016 \\
		\hline
	\end{tabular}
\end{table}

\begin{figure}[!hb]
	\centering
	\includegraphics[width=0.6\linewidth]{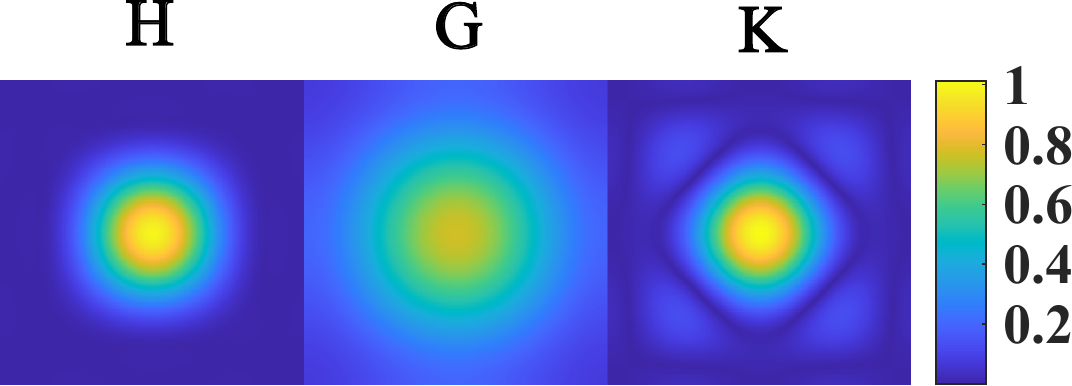}
	\caption{The spectrum of learned kernel $H$ (left) , $G$ (middle) and $K$ (right).}
	\label{fig:4}
\end{figure}

As mentioned, the learned kernels $H$ and $G$ are corresponding to the low-pass and high-pass filters in wavelet transform, respectively. Therefore, we analyze the spectrum of $H$, $G$ and $K$. The result is shown in Fig.~\ref{fig:4}. This result confirms that the learned $H$ and $G$ are indeed the low-pass filter (concentrated around the center) and the high-pass filter (covers more high frequency region), respectively. 


\subsubsection{Scene Adaptability}
We also investigate the scene adaptive ability of the proposed network. We use the well-known Mip-Nerf360 data set~\cite{Barron2022}, which contains multi scenes. Each scene has multi images (the number of images is shown in the second line in Table~\ref{tab:category_loss}). We train and test the network on each scene with $80\%$ images as training set and the rest as testing test, respectively. And the results are summarized in Table~\ref{tab:category_loss}. As shown in the bottom line, training across all the scenes has a satisfying generalization. Meanwhile, training on each scene, indeed, can improve the reconstruction accuracy.  

\begin{table}[htbp]
	\caption{Loss Comparison Across Different Scenes}
	\begin{center}
		\begin{tabular}{|c|cccccc|}
			\hline
			\rowcolor{gray!30}  \textbf{Train}$\backslash$Test& Stump & Bonsai & Counter & Kitchen & Room & Garden \\
			\rowcolor{gray!10} (Images)& (125) & (292) &(240) & (279)& (311) & (185) \\
			\hline
			\textbf{Stump} & \cellcolor{blue!10}1.225 & 1.658 & 2.131 & 3.420 & 2.224 & 3.557 \\
			\textbf{Bonsai} & 1.648 & \cellcolor{blue!10}1.272 & 2.037 & 3.235 & 2.159 & 3.662 \\
			\textbf{Counter} & 1.539 & 1.427 & \cellcolor{blue!10}1.501 & 3.318 & 2.112 & 3.486 \\
			\textbf{Kitchen} & 1.352 & 1.383 & 1.743 & \cellcolor{blue!10}2.154 & 1.814 & 3.033 \\
			\textbf{Room} & 1.408 & 1.427 & 1.802 & 2.729 & \cellcolor{blue!10}1.709 & 3.215 \\
			\textbf{Garden} & 1.796 & 1.435 & 2.504 & 4.052 & 2.575 & \cellcolor{blue!10}2.334 \\
			\hline
			\rowcolor{gray!20}\textbf{ALL} & 1.826 & 1.721 & 1.947 & 2.602 & 1.935 & 4.179 \\
			\hline
		\end{tabular}
		\label{tab:category_loss}
	\end{center}
\end{table}

\subsection{Compactness and Performance}
\subsubsection{Compactness} The proposed network has only 177 parameters and can be used to solver the image reconstruction task from its Laplacian field with any resolution. This leads to a very compact neural network. With such compactness, the proposed network can be deployed on most of hardware. 

\subsubsection{Accuracy} 
\begin{figure*}
	\centering
	\subfigure[original GT]{\includegraphics[width=0.18\linewidth]{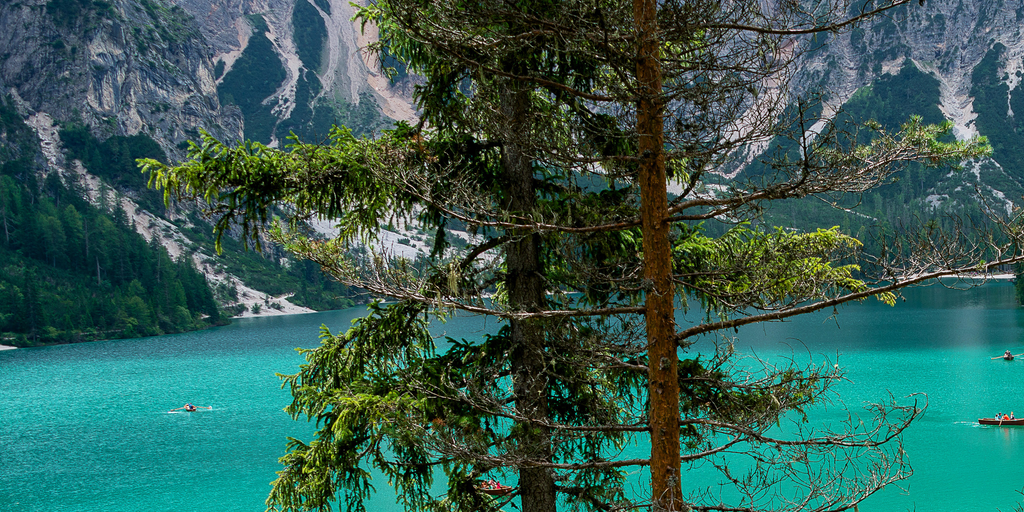}}
		\subfigure[wavelet, $MSE=17.39$]{\includegraphics[width=0.18\linewidth]{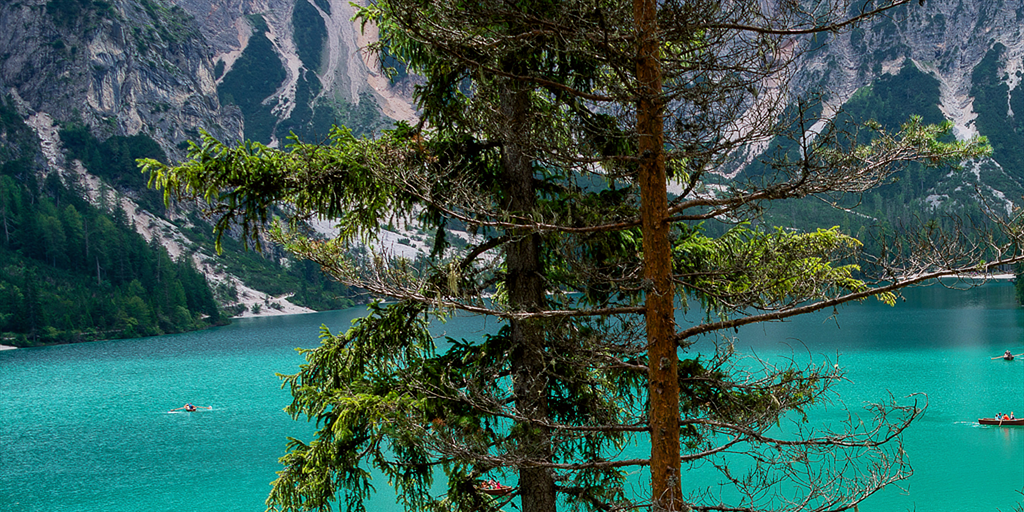}}
			\subfigure[wavelet error map]{\includegraphics[width=0.215\linewidth]{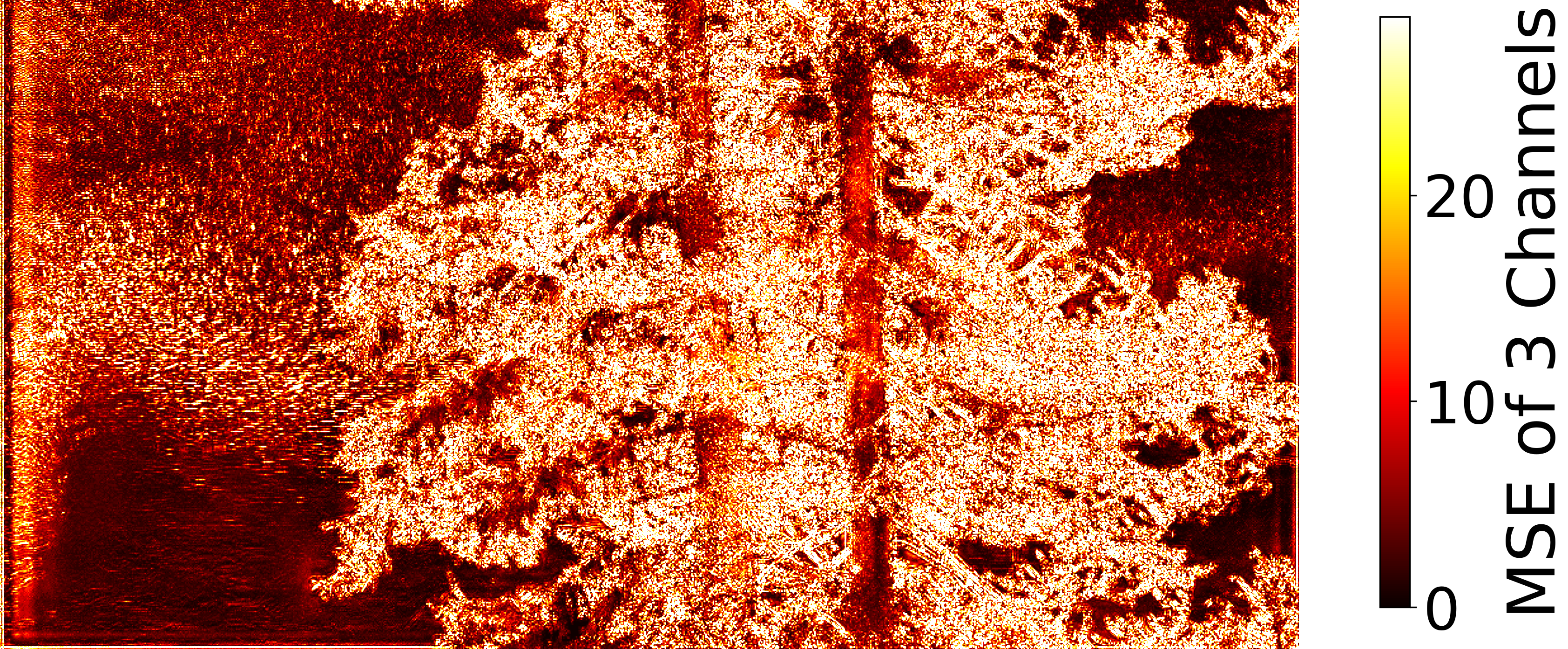}}
			\subfigure[ours, $MSE=1.37$]{\includegraphics[width=0.18\linewidth]{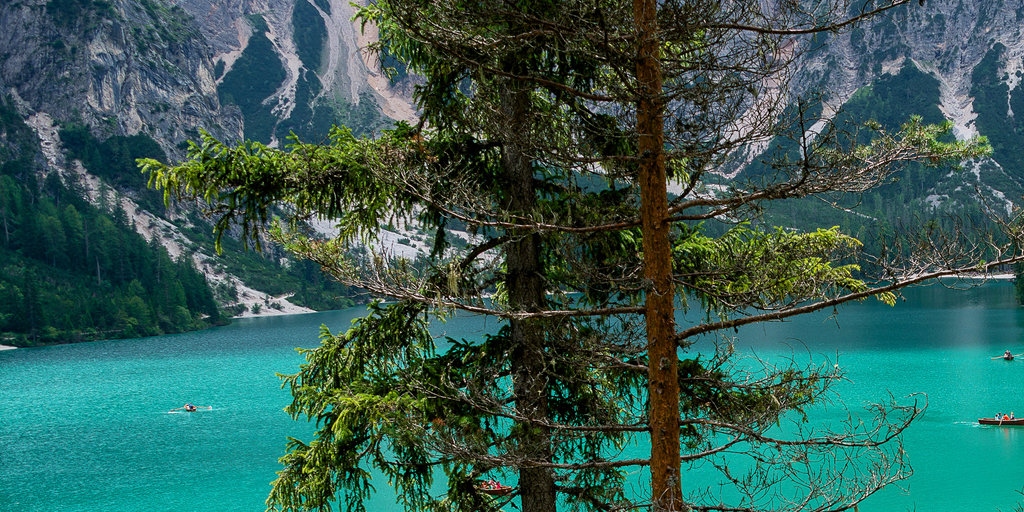}}
			\subfigure[our error map]{\includegraphics[width=0.215\linewidth]{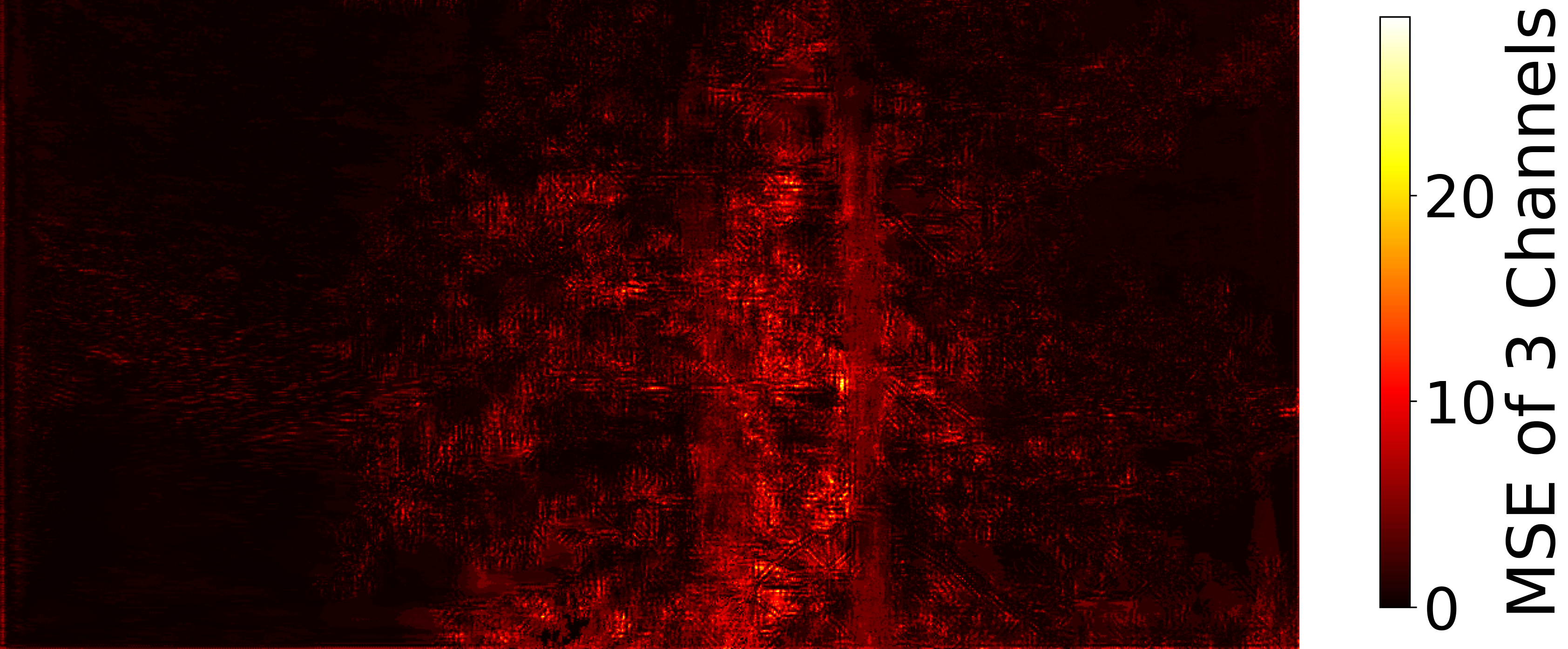}}
			
				\subfigure[original GT]{\includegraphics[width=0.18\linewidth]{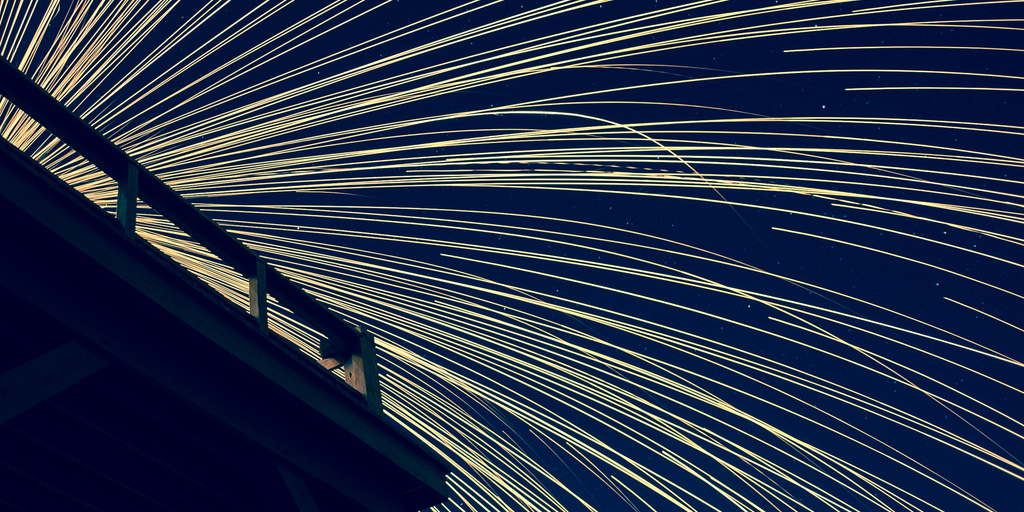}}
			\subfigure[wavelet, $MSE=16.12$]{\includegraphics[width=0.18\linewidth]{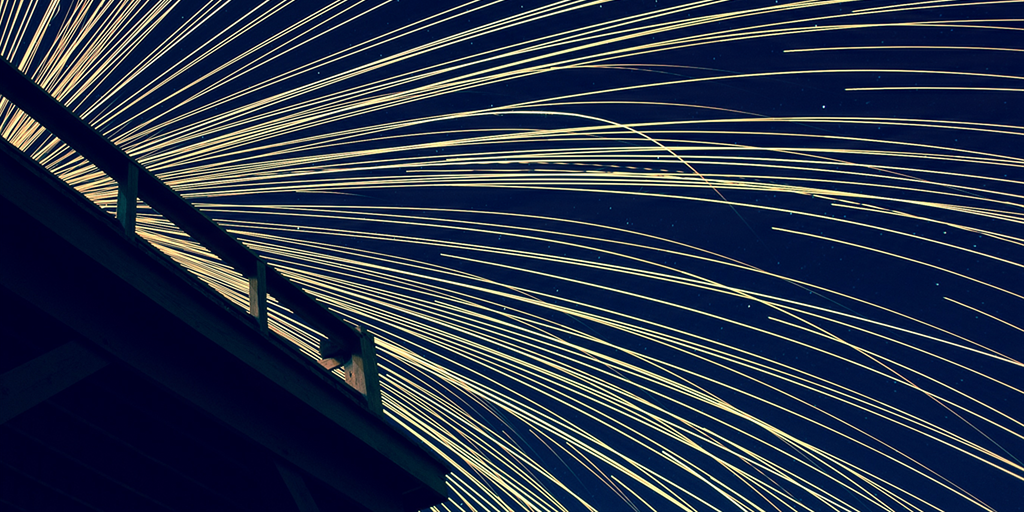}}
			\subfigure[wavelet error map]{\includegraphics[width=0.215\linewidth]{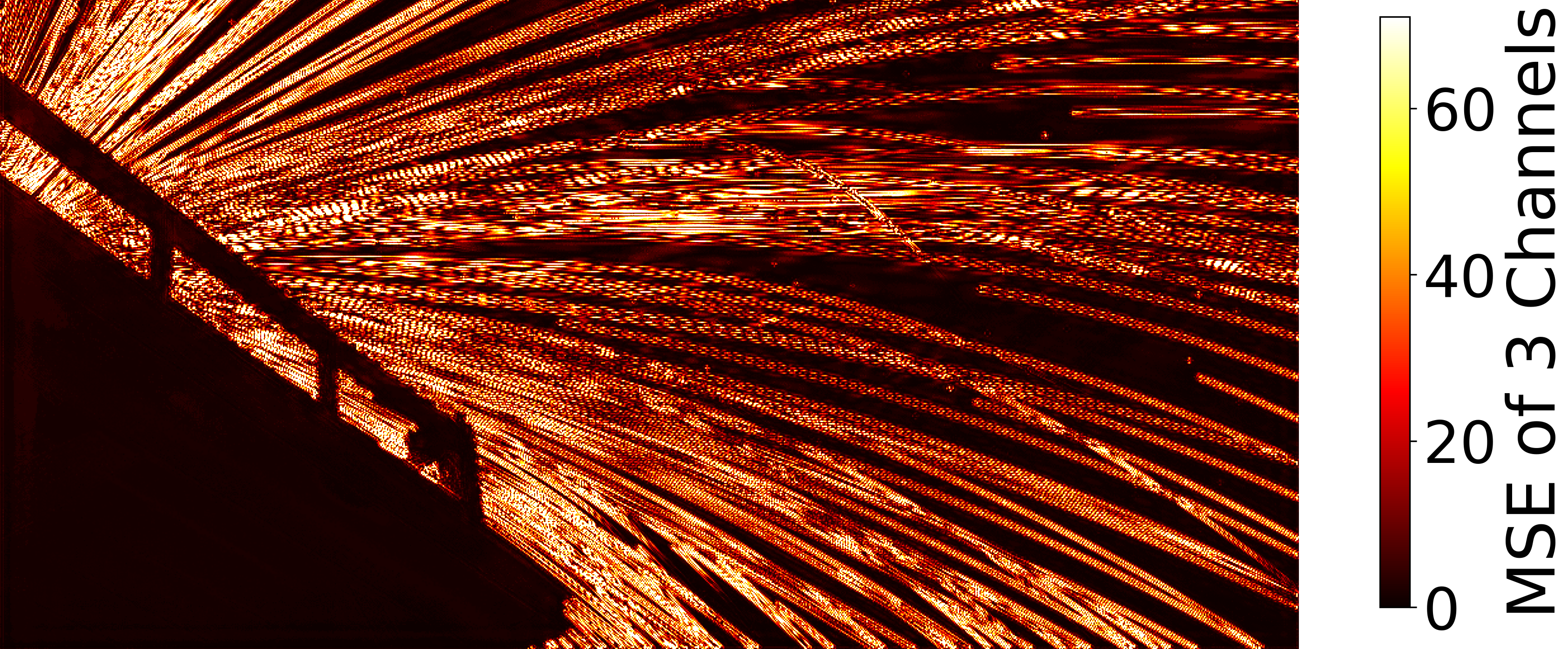}}
			\subfigure[ours, $MSE=1.36$]{\includegraphics[width=0.18\linewidth]{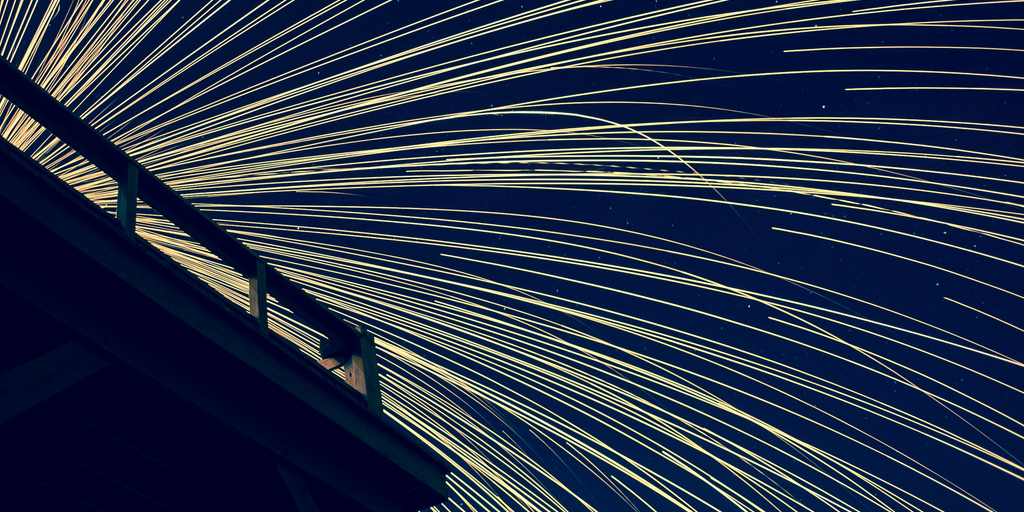}}
			\subfigure[ours error map]{\includegraphics[width=0.215\linewidth]{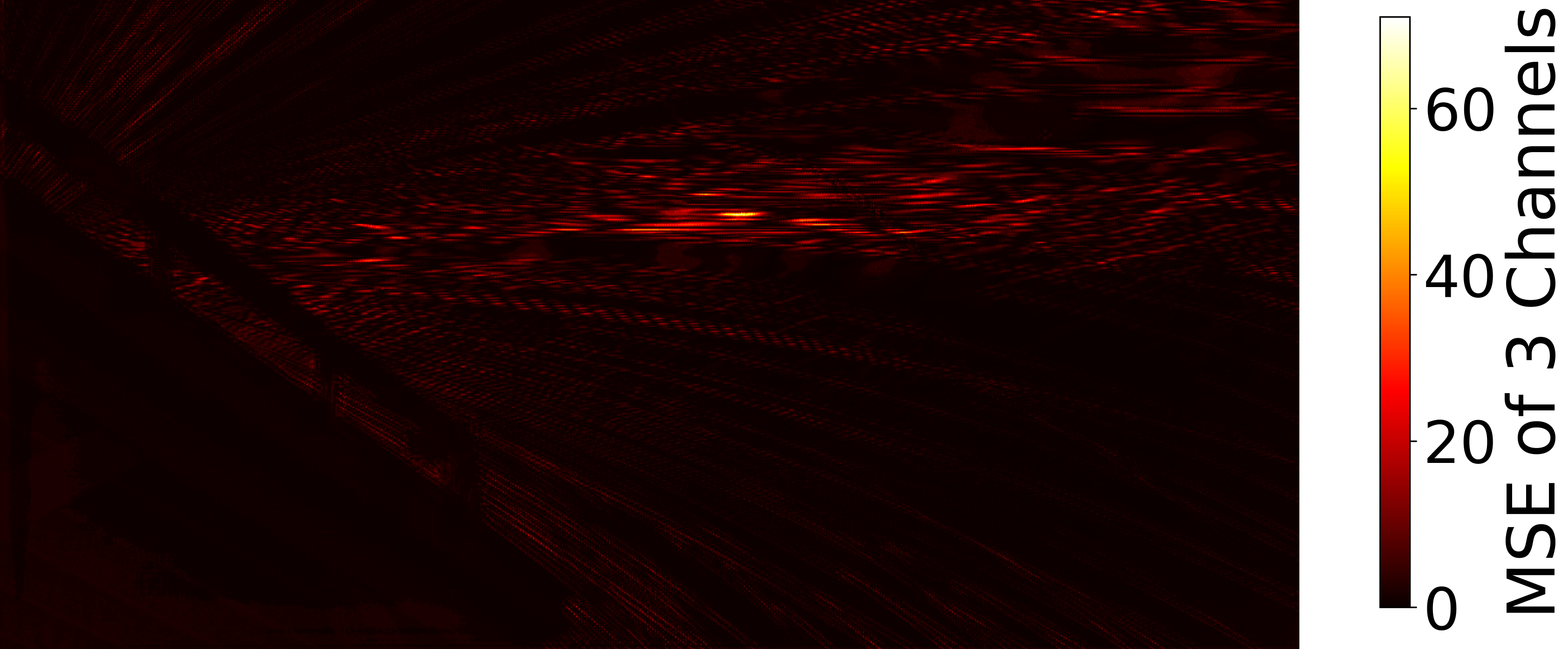}}
	\caption{From left to right: original (a, f), wavelet reconstruction (b, g), the reconstruction error (c, h), our reconstruction (d, i) and our error (e, j). For the error map, (c) and (e) share the same color range. (h) and (j) share the same color range. These results confirm the higher accuracy of our method.}
	\label{fig:exp}
\end{figure*}

The proposed network is also accurate than the previous wavelet method~\cite{Farbman2011}, as numerically confirmed in Table~\ref{tab:loss_comparison}. Two examples are visually compared  in Fig.~\ref{fig:exp}.

\subsubsection{High Performance} Despite the accuracy, the proposed network has linear computational complexity, leading to a high computational performance. We compared the method with other solvers in Table~\ref{tab:time} and the results confirm our conclusion.
\begin{table}[htbp]
	\caption{Our method is fast as the wavelet method.}
	\label{tab:time}
	\centering
	\begin{tabular}{|c|c|c|c|c|}
	\hline
	Method$\backslash$resolution& 256*256 & 512*512 & 1024*1024 & 2048*2048 \\
	\hline
	DCT & 6.7279 & 3.9335 & 22.3457 & 307.0218 \\
	\hline
	wavelet & 1.3085 & 1.5241 & 1.5191 & 2.6849 \\
	\hline
	ours & 1.0934 & 1.2238 & 1.4664 & 4.8238 \\
	\hline
\end{tabular}
\end{table}
\if false
\begin{figure}[!hbt]
	\centering
	\includegraphics[width=0.5\linewidth]{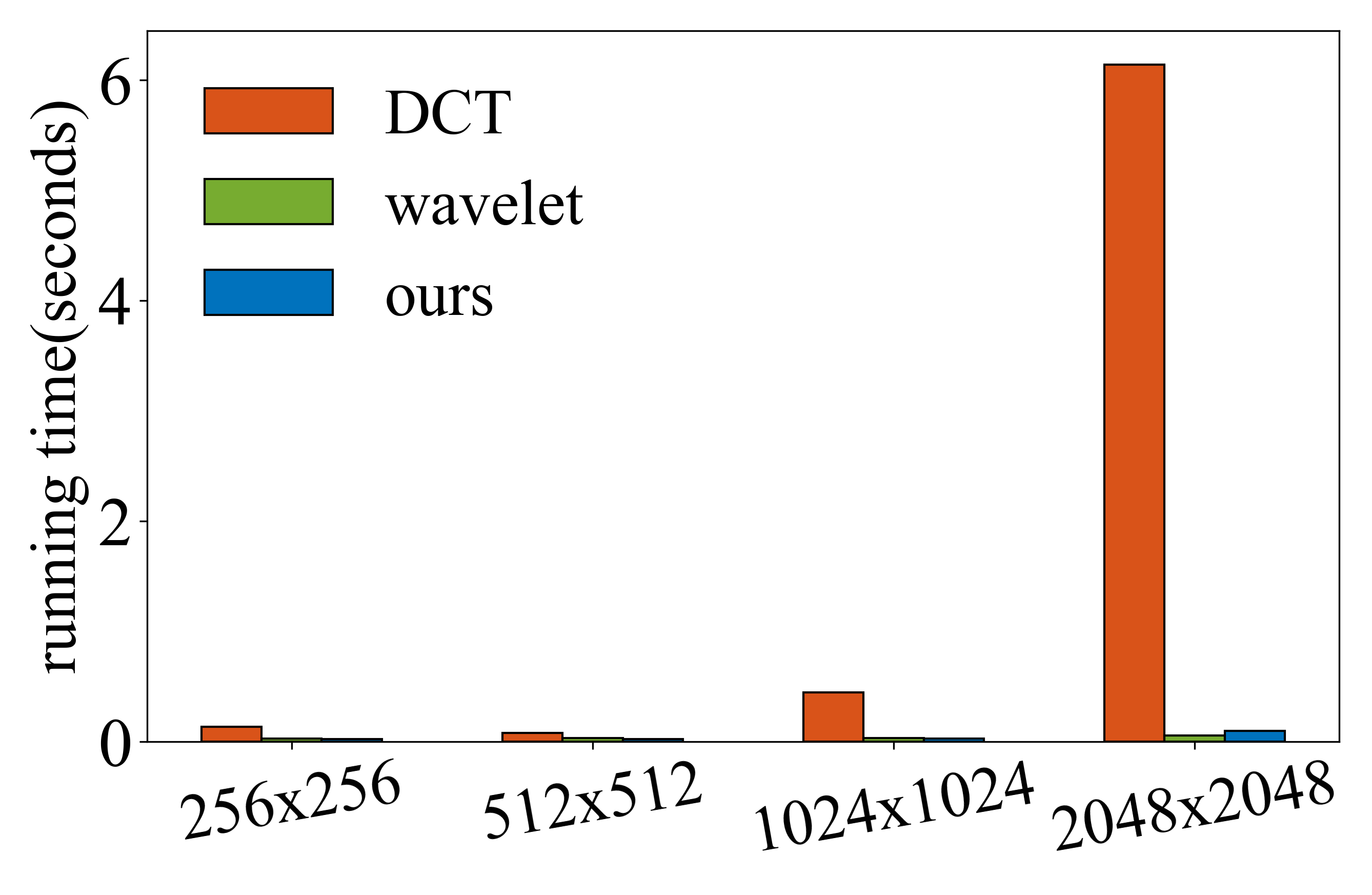}
	\caption{Our method is fast as the wavelet method and much faster than DCT.}
	\label{fig:time}
\end{figure}
\fi

\section{Conclusion}
In this paper, we propose to represent an image by its Laplacian field, which is sparse and satisfies a stable distribution. Moreover, the original image can be accurately and efficiently reconstructed from its Laplacian field via solving a Poisson equation. To achieve this task, we proposed a novel neural network solver that has several advantages. First, the proposed network inherits the wavelet transformation. Second, the convolution kernels are shared crossing different scales, leading to a compact neural network with only $0.0002M$ parameters. Several numerical experiments are conducted to confirm the effectiveness and efficiency of the proposed neural network solver. The proposed method can be applied in a large range of applications~\cite{Gong2017}, such as fluid simulation~\cite{Sousa2024}, view synthesis~\cite{Gong2024}, 3D surface reconstruction, motion deblurring~\cite{Gong2024a} and image approximation~\cite{Gong2021,Gong2024b,Gong2025}.

\bibliographystyle{IEEEbib}
\bibliography{icme2025references}

\end{document}